\documentclass[11pt,twoside]{article}

\usepackage{asp2006}
\usepackage{epsf}
\usepackage{lscape}

\newcommand{\msun}{M$_\odot$}

\newcommand{\bb}{$b_{435}$}
\newcommand{\vv}{$v_{606}$}
\newcommand{\ii}{$i'_{775}$}

\begin{document}
\title{The Star Formation Density at z$\sim$7}   %%% Fill in title
\author{F. Mannucci}   %%% Fill in author names
\affil{INAF - IRA, Firenze}    %%% Fill in author affiliations

\begin{abstract} %%% Abstract to run on from here.
Near-infrared VLT data of the GOODS-South area 
were used to look for galaxies at z$\sim$7
down to a limiting magnitude of (J+Ks)$_{AB}$=25.5.
No high-redshift candidates were detected, and this
provides
clear evidence for a strong evolution of the luminosity function
between z=6 and z=7, i.e. over a time interval of only $\sim$170~Myr.
Our constraints provide
evidence of a significant decline in the total star formation rate
at z=7, which must be less than 40\% of that
at z=3 and 40-80\% of that at z=6.
The resulting upper limit to the
ionizing flux at z=7 is only marginally
consistent with what is required to completely ionize the Universe.
\end{abstract}

%---------------------------------------------------------------------------

\section{Introduction}
\label{sec:intro}

In recent years, a large effort was put into 
obtaining a clear picture of the evolution of the star formation activity
along the Hubble time.
Observations are approaching the interesting redshift range between z=6 and
z=10 where current cosmological models expect to find the
end of the reionization period and
the ``starting point'' of galaxy evolution (e.g., Stiavelli,
Fall \& Panagia 2004).

Most high-redshift galaxies 
were selected by the {\em dropout} technique, i.e., by
detecting the spectral break in the UV continuum blueward of the Ly$\alpha$\
due to intervening Ly$\alpha$-forest.
This technique is mainly sensitive to high-redshift
UV-bright galaxies, commonly named Lyman-Break Galaxies (LBGs). 
The use of different filters allows
the selection of different redshift ranges.
At z$>$6 the use of near-IR images is mandatory,
which makes detection much more difficult. 
Deep J-band images from HST/NICMOS exist, but their
field-of-view is limited to a few sq.arcmin. 
Larger fields can be observed 
by ground-based telescopes, but at the expense of 
lower spatial resolution and brighter detection limits.
 
Recently, Bouwens \& Illingworth (2006) looked for galaxies
at z$\sim$7 in a few deep NICMOS fields covering
$\sim$19 sq.arcmin.
They detected  four possible z$\sim$7 objects, while 17 were expected
based on the z=6 LF. 
This result point toward the existence of a
strong reduction in the LF with increasing redshift at z$>$6.
In contrast, Richard et al. (2006) examined two lensing clusters and 
derived a star formation density (SFD)
well in excess of the one at z=3, hinting at large amounts of star formation 
activity during the first Gyr of the universe. 
The faintness of the high-redshift candidates of the previous studies
imply that their redshifts cannot be spectroscopically confirmed
with the current generation of telescopes.
Here we present a study aimed at detecting
brighter z=7 objects in a large field,
in order to measure the cosmic star-formation density at this redshift.

%----------------------------------------------------------------------

\begin{figure}[t]     
\plotfiddle{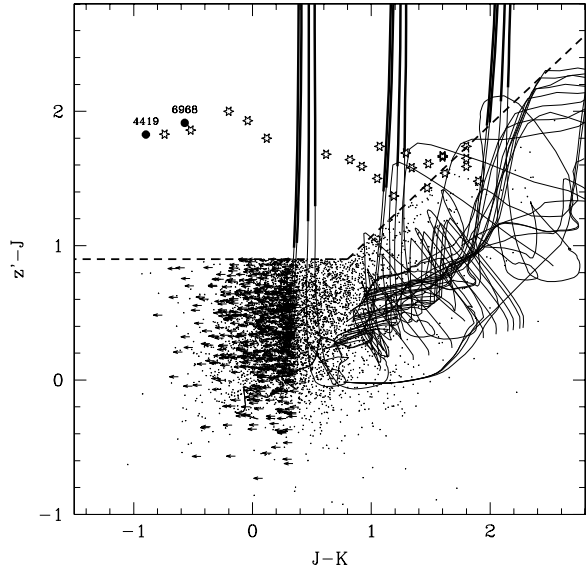}{6.8cm}{0}{40}{40}{-100}{-70}
\caption{
\label{fig:newsel}
Color-color selection diagram for z$>$7 galaxies. 
The solid lines show the variation in the colors with redshift
expected for local galaxies (Mannucci et al. 2001) 
and galaxy models by the Bruzual \& Charlot (2003)
with three different amounts of extinctions (E(B-V)=0.0, 0.3, and 0.6).
Thin and thick lines show the expected colors for galaxies below and above
z=7, respectively.
Stars show the expected positions of Galactic brown dwarfs
(ranging from  T8 to L1, from left to right). 
Small dots show the galaxies detected in the GOODS-South in all three
bands, while the arrows show the positions of the objects undetected in K. 
The dashed line shows the color threshold. Above this threshold only objects
with no counterparts either in the HST/ACS filters \bb, \vv, and \ii\ 
or in the sum \bb+\vv+\ii\ image are shown.
Large solid dots with numeric label 
show the measured colors of two brown dwarfs.
}
\end{figure} 

%----------------------------------------------------------------------
\section{Observations, object catalog and color selection} 
\label{sec:obs}

The main catalog is based on ESO/VLT J and Ks data
of the GOODS-South field covering 133 sq.arcmin.
The typical exposure time is 3.5h in J and 6h in K, with a seeing of about 
0.45\arcsec\ FWHM.
To improve detectability of blue objects we used the 
J+Ks sum image to build the main object catalog.
The average 6$\sigma$ AB magnitude limit inside a 1\arcsec\ 
aperture is 25.5, estimated 
from the statistics of the sky noise. 
The main catalog comprises about 11.000 objects.

We selected {\em drop-out} objects that are undetected ($<1\sigma$) 
in \ii\ and in all bluer bands and that have $z'$--J$>$0.9 
and $\rm{J-Ks}<1.2(z'-\rm{J})-0.28$ as measured in 1\arcsec\ apertures, 
as shown in Figure~\ref{fig:newsel}. 
It is evident that
the use of near-IR colors alone is enough to exclude
low/intermediate galaxies, but not to distinguish true z$\sim$7 star-forming
galaxies from high-z QSOs or Galactic brown dwarfs. 

Two objects show colors that are
compatible with star-forming galaxies at z$>$7. 
Through detailed analysis of their morphology and
near/mid-IR colors, as well as through spectroscopic information,
these two candidates were identified as Galactic brown dwarfs
(for details, see Mannucci et al., 2006)
As a consequence, no z=7
object is present in the survey field above our detection limit.

%--------------------------------------------------------------
\section{Evolution of the luminosity function and of the cosmic 
star-formation density} 
\label{sec:lf}

The detection of no z$>$7 objects in the field can be used to place an upper
limit to the LF of these objects. 
To do this we have computed the effective sampled volume as a function
of the absolute magnitude of the objects taking into account the effect of
redshift on both colors and brightness of the objects.
We use the concordance cosmology 
$(h_{100},\Omega_m,\Omega_{\Lambda})=(0.7,0.3,0.7)$.
The redshift sensitivity of our selection method starts at about z=6.7, 
followed by a peak at z=7 and a shallow decrease towards high redshifts. 
The limiting apparent magnitude ($\sim$25.5) corresponds, at z=7, to
an absolute magnitude M(1500)=--21.4, i.e., to
SFR$\sim$20\msun/yr (Madau et al. 1998). This means that we can sample
the bright part of the LF of LBG at any redshift down to $\sim$ 1--2 $L^*$.

In the left panel of figure~2 we 
show the 1$\sigma$ upper limits to the density
of objects, compared with the LF of the LBGs 
at z=6 from Bouwens et al. (2006) 
and at z=3 obtained by Steidel et al. (2003).
Our upper limits imply 
an evolution of the LF from z=6, even if only 170 Myr have passed since then. 
In the no-evolution case, in fact,
we would expect to detect 5.5 objects, 
with a Poisson probability of no detection smaller than 0.5\%.

Using a confidence level CL=90\% and assuming density 
evolution of the LF, we found that the normalization of the LF 
at z=7 must be at most 40\% of that at z=6.
Assuming luminosity evolution we obtain that the objects 
must be at least 0.22 mag brighter at z=6 than at z=7.
This evolution corresponds to a reduction 
of 20\% in the total SFD, obtained by integrating the LF 
assuming a constant faint-end slope $\alpha$.

Our limits can be compared with the LF at z=3 (Steidel et al. 2003).
Assuming luminosity evolution
we obtain a shift of $L^*$ about 0.9 mag, corresponding to 
SFD(z=7)/SFD(z=3)=0.42, a reduction of more than
a factor of 2.  
A pure density evolution would require SFD(z=7)/SFD(z=3)=0.05.

%--------------------------------------------------------------
\begin{figure}[ht]  
\plotfiddle{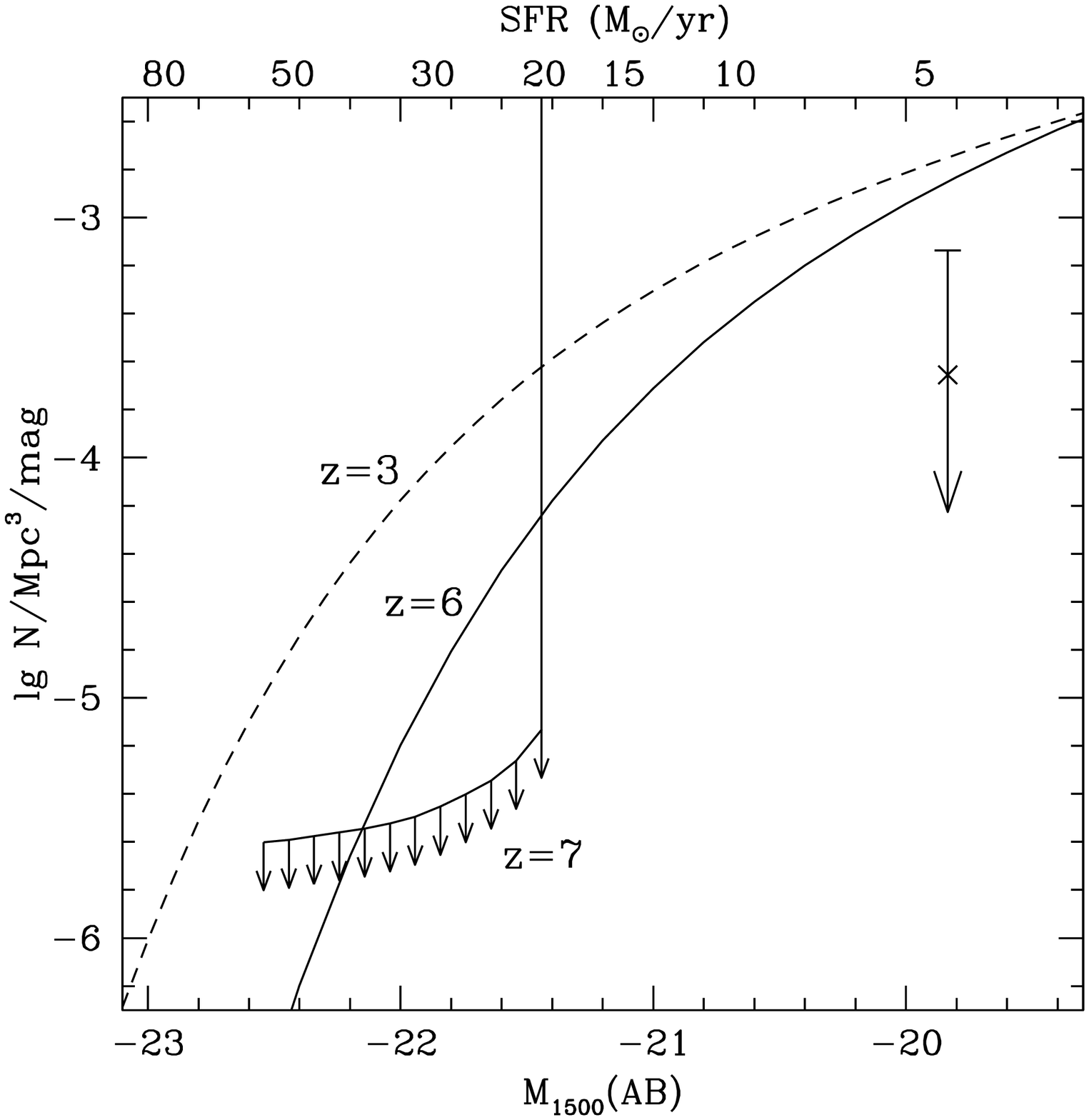}{7.0cm}{0}{32}{32}{-200}{-30}
\plotfiddle{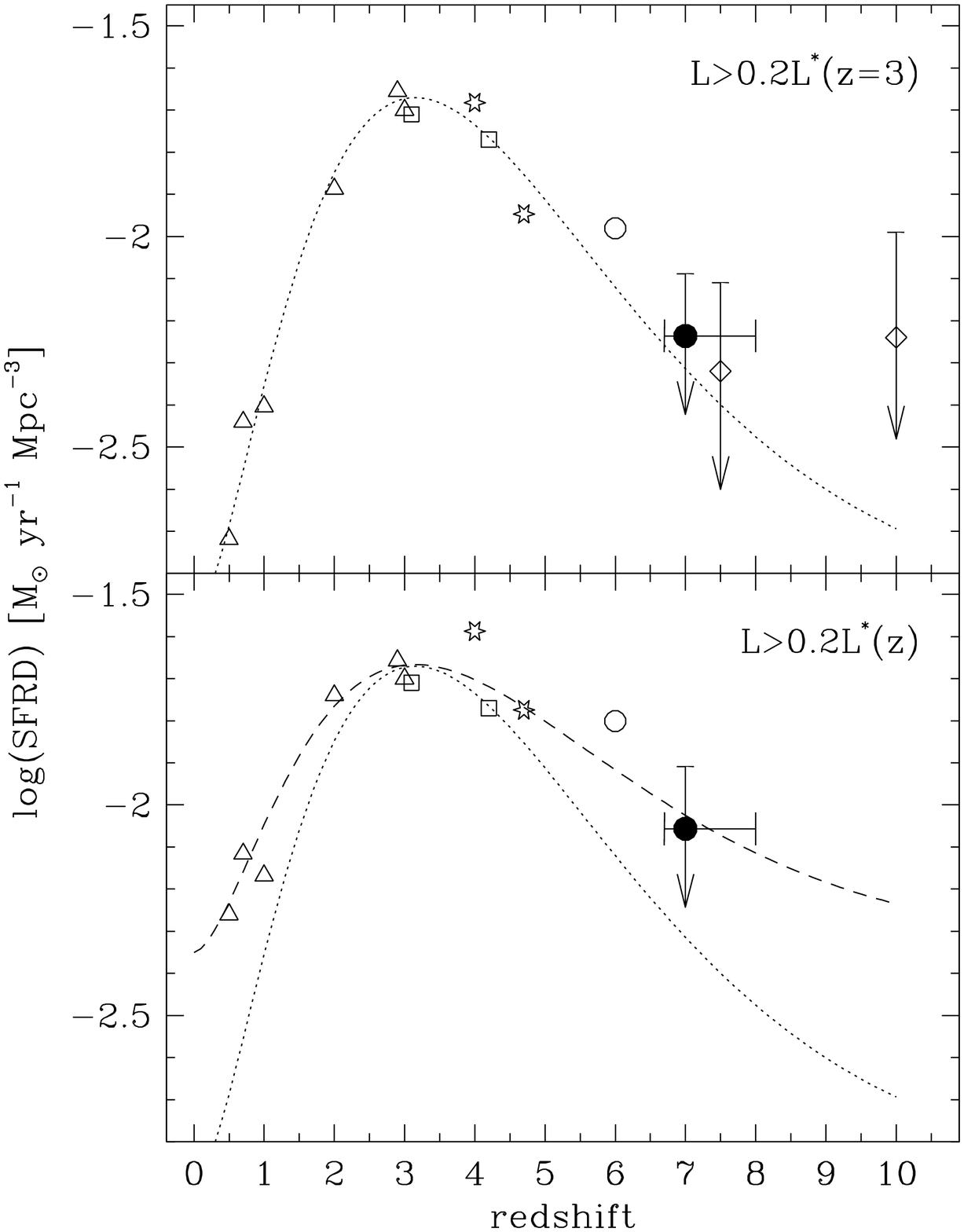}{0.1cm}{0}{35}{35}{-10}{-20}
\caption{
\label{fig:lf}
{\em Left:} 
The limits to the density of z=7 galaxies (connected downward arrows)
are compared with the LFs of LBGs at z=3 (dashed line, Steidel et al. 2003) 
and z=6 (solid line, Bouwens et al. 2006)
The upper limit with a cross corresponds to the object detected by
Bouwens et al. (2004) in the HUDF and considered to be at z$\sim$7.
\newline
{\em Right:}
Cosmic star formation as derived from 
rest-frame UV observations and with no correction for dust extinction.
The upper panel shows the results of integrating the LF for $L>0.2L^*(z=3)$,
the lower panel for $L>0.2L^*(z)$ (see text).
The black dot with error bars is
the upper limit from the present work and is the average
the average between the two extremes of
pure luminosity and pure density evolution. 
while the upper value corresponds
to pure luminosity evolution.
The empty dots are from literature
(see Mannucci et al., 2006, for details).
The diamonds at z=7.5 and z=10 are from Bouwens et al. (2004, 2005).
The dotted lines shows an empirical fit to the data in the upper panel,
the dashed line to those in the lower panel.
}
\end{figure}  %
%--------------------------------------------------------------

%--------------------------------------------------------------------
\section{The evolution of the star formation activity}
\label{sec:sfd}

In the right panel of figure~2 we show the values of the SFD obtained 
by integrating the observed
LFs above a given luminosity threshold and converting to SFR using Madau et
al. (1998). 
The most interesting quantity, the
total SFD at each redshift, would be obtained by using a very low luminosity
threshold. 
Unfortunately, this would introduce large uncertainties due to the
unobserved part of the LF at low luminosities. For example, 
for a faint-end slope of the LF $\alpha=-1.6$, about half of the
total SFD takes place in galaxies below 0.1$L^*$, which are usually unobserved.
This implies that a
correction of about a factor of two is needed to obtain the total SFD from the
observed part of the LF. 
To avoid this additional uncertainty, it
is common to refer to the SFD derived from galaxies that where 
directly observed or by
using a small extrapolation of the LF.

The right panels of figure~2 show the SFD as
obtained by integrating the observed LFs with two different lower limits. 
In the {\em upper panel} we show the results 
when considering, at any redshift,
the same range of absolute magnitudes ($M(1500)>$-19.32), i.e., 
limiting the luminosity to above 20\% of the $L^*$ magnitude
derived by Steidel et al. (1999) for their LBGs at z=3
($L>0.2L^*(z=3)$). 
The derived SFD is directly related to the total SFD in the case of 
pure density evolution, as the fraction of SFD from galaxies below 
the threshold would be constant. 
The use of the absolute limits of integration is 
very common (see, for example, Bouwens et al. 2006)
The resulting SFD appears to vary rapidly, increasing by more than a 
factor of 30 from z=0.3 to z=3 and then decreasing by a 
factor of 2.2 to z=6 and $\sim$4 to z=7.

In the case of luminosity evolution, the use of a constant 
limit of integration results in considering a variable fraction of LF.
For example, Arnouts et al. (2005) measured the UV LF at low redshift
and found a strong luminosity evolution, with $M^*$=--18.05 at z=0.055
and $M^*$=--20.11 at z=1.0. In this case, at low redshift
the limit of integration used above ($M(1500)>$--19.32) 
is even brighter than $L^*$ and only a small fraction of
the LF is integrated to obtain a value of the SFD. This is the
reason that the upper
panel of figure~2 shows such a strong evolution at low redshift.
In the {\em lower panel}  of figure~2
we use a variable
limit of integration, set to be $0.2L^*(z)$ at each redshift. This is more
suitable to reproducing the total cosmic SFD, as the luminosity evolution
appears to dominate both at low (Arnouts et al. 2005) and at high
redshifts (Bouwens et al. 2006). In this case, the obtained evolution is much
milder with an increase of a factor of 5 from z=0.3 to z=3 and a decrease
of $\sim$1.4 to z=6 and $\sim$2.5 to z=7. 
The resulting values of the UV luminosity density and 
of the SFD can be found in Mannucci et al. (2006).

\smallskip

All the data in figure~2 are derived from UV observations
and, as a consequence, they are very sensitive to dust extinction,
as discussed by a large number of authors.
Variation in the dust content along the cosmic age is one of the effects
that could contribute to shaping the observed evolution of the SFD. 
The typical color of z=6 LBGs, represented by the UV spectral slope $\beta$,
is bluer at z=6 
($\beta=-2.2$, Stanway et al. 2005) than at z=3 
($\beta=-1.5$, Adelberger \& Steidel 2000), pointing towards a
reduction in average extinction at high redshift of about a factor of two. 
As a consequence, the observed reduction in the SFD cannot be due to
an increase of the dust content. Actually, considering this effect 
would make the increase of SFD with cosmic time even more pronounced.

\smallskip

As we observe the bright part of the LF, we cannot exclude 
that the reduction in the number density of bright galaxies is 
not associated with an increase in that of the faint galaxies.
If, for
example, both $\Phi^*$ and $L^*$ vary together so that $\Phi^*L^*$ remains
constant, the resulting total SFD also remains constant. We cannot exclude
this effect, even if the upper limits to the galaxies with $L\sim L^*$
from Bouwens et al. (2004) tend to exclude this possibility.

\section{Consequences on reionization of the primordial universe}

It is widely accepted that high redshift starburst galaxies can 
contribute substantially to the reionization of the universe.
Madau et al. (1999)  estimated the amount of 
star formation needed to provide enough ionizing photons to the intergalactic
medium. By assuming an escape fraction of the photons $f_{esc}$ of 0.5 and a
clumping factor $C$ of 30 (Madau et al. 1999), 
we find that at z=7 the necessary SFD is

$$ \rm{SFD(needed)}\sim7.8\times10^{-2}~~~\rm{M}_\odot\rm{yr}^{-1}\rm{Mpc}^{-3} $$

\noindent
The observed total amount of SFD derived from UV observations can be derived 
by integrating the observed LF down to zero luminosity. 
Assuming luminosity evolution and integrating the LF down to $0.01L^*$,
we obtain 

$$\rm{SFD(observed)}=2.9\times10^{-2}~~~\rm{M}_\odot\rm{yr}^{-1}\rm{Mpc}^{-3}$$

\noindent
where about half of this comes in very faint systems, below 0.1$L^*$.
This value can be increased up to 
$\sim5\times10^{-2}\rm{M}_\odot/\rm{yr}^{-1}\rm{Mpc}^{-3}$
assuming the 
steeper faint-end slope on the LF ($\alpha=-1.9$) compatible with the data
in Bouwens et al. (2006). 

The uncertainties involved in this computation 
are numerous and large for both SFD(needed) and SFD(observed).
Nevertheless, the amount of ionizing
photons that can be inferred at z=7 from observations is less or, at most,
similar to the needed value. 
A measure of the SFD at even higher redshifts or tighter constraints
to the faint-end slope of the LF at z=6 can significantly reduce these
uncertainties and can 
easily reveal that it falls short of the required value.

\section{Conclusions}

The existing multi-wavelength deep data on the large 
GOODS-South field allowed us to search for z=7 star-forming galaxies
by selecting $z'$-dropouts.
The accurate study of the dropouts in terms of colors, morphology, and spectra
allows us to exclude the presence of any z=7 galaxy in the field above 
our detection threshold. We used this to derive evidence for the evolution 
of the
LF from z=7 to z=6 to z=3, and to determine an upper limit to the global 
star formation density at z=7.
These limits, together with the numerous works at lower redshifts,
point toward the existence of a sharp increase of the star 
formation density with cosmic time from z=7 to z=4, a flattening
between z=4 and z=1, and a decrease afterward.
The ionizing flux from starburst galaxies at z=7 could be too low to 
produce all the reionization.


\begin{thebibliography}{}

\bibitem{}
  Adelberger, K. L., \& Steidel, C. C., 2000, ApJ, 544, 218				%OK
\bibitem{}
  Arnouts, S., Shiminovich, D., Ilbert, O., et al., 2005, ApJ, 619, L43 %OK
\bibitem{} 
   Bouwens, R. J., Thompson, R. I., Illingworth, G. D, et al., 
   2004, ApJ 616, 79 		% z-drop                                    %OK
\bibitem{} 
   Bouwens R. J., Illingworth, G. D., Thompson, R. I., \& Franx, M., 
   2005, ApJ 642, 5  		% J-drop                                   %OK
\bibitem{} 
   Bouwens, R. J., \& Illingworth, G. D., 2006, Nature, in press
   (astro-ph/0607087) %OK
\bibitem{} 
   Bouwens, R. J., et al., 2006, in press (astroph/0509641)    % i-drop, LF                    %OK
\bibitem{}
  Bruzual, G. \& Charlot, S., 2003, MNRAS, 344, 1000 	               %OK
\bibitem{}
  Madau, P., et al., 1998, ApJ 498, 106 
\bibitem{}
  Madau, P., Haardt, F., Rees, M., 1999, ApJ, 266, 713					%OK
\bibitem{}
 Mannucci, F., et al., 2001, MNRAS, 326, 745
\bibitem{}
 Mannucci, F., et al., 2006, in press (astro-ph/0607143)
\bibitem{}
  Richard, J., et al., 2006, in press (astro-ph/0606134)
\bibitem{}
  Stanway, E. R., McMahon, R. G., \& Bunker, A. J., 2005, MNRAS, 359, 1184 %OK
\bibitem{}
  Steidel, C. C., et al., 1999, ApJ, 519, 1  
\bibitem{} 
  Steidel, C. C., et al., 2003, ApJ, 592, 728
\bibitem{}
  Stiavelli, M., Fall M., \& Panagia, N., 2004, ApJ, 610, 1

\end{thebibliography}
\end{document}